# Bacterial morphologies in carbonaceous meteorites and comet dust

Chandra Wickramasinghe[a]*, Max K. Wallis[a], Carl H. Gibson[b], Jamie Wallis[a], Shirwan Al-Mufti[a] and Nori Miyake[a]

[a] Cardiff Centre for Astrobiology, Cardiff University, UK.
[b] Depts of Mechanical and Aerospace Engineering and Scripps Institution of Oceanography, Center for Astrophysics and Space Sciences, University of California at San Diego, La Jolla CA 92093-0411, USA

*Corresponding author: Email: ncwick@gmail.com, Wickramasinghe@cf.ac.uk

## ABSTRACT

Three decades ago the first convincing evidence of microbial fossils in carbonaceous chondrites was discovered and reported by Hans Dieter Pflug and his collaborators. In addition to morphology, other data, notably laser mass spectroscopy, confirmed the identification of such structures as putative bacterial fossils. Balloon-borne cryosampling of the stratosphere enables recovery of fragile cometary dust aggregates with their structure and carbonaceous matter largely intact. SEM studies of texture and morphology of particles in the Cardiff collection, together with EDX identifications, show two main types of putative bio-fossils – firstly organic-walled hollow spheres around 10μm across, secondly siliceous diatom skeletons similar to those found in carbonaceous chondrites and terrestrial sedimentary rocks and termed 'acritarchs'. Since carbonaceous chondrites (particularly Type 1 chondrites) are thought to be extinct comets the data reviewed in this article provide strong support for theories of cometary panspermia.

## 1. INTRODUCTION

Solar system cometary bodies endowed with radioactive heat sources (Wallis 1980, Wickramasinghe et al, 2009) provided a site for the replication of micro-organisms accreted from the interstellar dust cloud. Earlier generations of similar comets were similarly potential sites for *ab initio* origin of life at some earlier stage (Napier et al, 2007; Gibson, Schild and Wickramasinghe, 2010). Within an individual cometary body endowed with nutrients and chemical energy, pre-existing microbiota can proliferate on a very short timescale (Hoyle and Wickramasinghe, 1981, 1982) and undergo evolution over the Myr duration of a vapour-liquid interior. Thereafter the amplified microbial population becomes locked in a frozen state until the comet comes to be peeled away layer by layer, thus releasing viable microbes into space (Hoyle and Wickramasinghe, 1985). Along with the comet's loss of volatiles over many perihelion passages, a crust of mineral and carbonaceous particles builds up, with a periodically warmed layer below. We have shown sub-surface lakes would form, as evidenced in ice-features on the surfaces imaged by recent comet probes (Wickramasinghe et al 2009). Biology is envisaged as reviving and developing in such sub-surface lakes and adjacent warmed ice, but the gradual loss of vapour eventually leaves sediments of mineral grains along with residual microorganisms, including fossil material of the original interior microbes. On this basis it is possible to understand an origin of type I carbonaceous chondrites such as the Murchison and Orgueil meteorite as products of cometary geophysics and bioprocessing.

The amazingly rich diversity of organics identified in the Murchison meteorite (Cronin et al. 1988; Schmitt-Koplin et al, 2010) comes as no surprise in the context of cometary panspermia. If cometary bodies are carriers of microbial life, a diversity of organic molecules richer than that which prevails on Earth is to be expected, generated more via biological than abiotic or prebiotic processes (Wickramasinghe, 2010). We conceive that extraterrestrial biology carried by comets with diverse initial complements and histories could represent a greater variety than the subset adapted to terrestrial environments and surviving competitive evolution. It is unnecessary to assign such a diversity to prebiotic processes as suggested by Schmitt-Koplin. For example, the exotic "non-biological" amino acids that are anomalously abundant for 100kyr across the K/T boundary – ie. AIB and isovaline – appear to indicate a novel but temporary addition to biology. It has been argued that the genes coding for peptides containing these aminoacids could have derived from a comet (Wallis 2007), consistent with the picture of a fragmenting giant comet producing the impactor that caused the K/T crater and iridium layer. From this viewpoint, the complex carbon compounds in the Murchison meteorite that include ~60 non-biological aminoacids, could be interpreted as degradation products of non-terrestrial biology – closer to early terrestrial biology prior to specialisation into the present genetic code (Vetsigian et al. 2006).

## 2. EARLY CLAIMS OF FOSSILS IN METEORITES

Early in the 1960's, Claus and Nagy (1961) identified possible microfossils in carbonaceous chondrites (CCs), supported by chemical bio-markers. Critics mounted a wide array of objections to discredit these discoveries, claiming, for example, that some fossils had been contaminated by ragweed pollen.

Methods for isolating fossil carbonaceous material from sedimentary rocks - dissolving the minerals with acids – were applied to the Orgueil meteorite by Rossignol-Strick and Barghoorn (1971). They discovered hollow spherical shells, which in terrestrial rocks would have presumed biological origin, but, according to the authors, could result from carbonaceous deposits on mineral particles.

Chemical studies have shown meteoritic carbonaceous material to be highly complex (analogous to kerogen or sporopollenin; Brooks and Shaw, 1969). A range of extractable organic compounds, including alkanes, alkenes, aminoacids and nitrogen heterocyclics, extending to the purine and pyrimidine bases of DNA, have been reported (Hayatsu and Anders 1981). However, these were presumed to derive abiotically from dust and condensing gases processed in interstellar space by stellar UV and other radiation – a process that we ourselves consider less likely (Wickramasinghe, 2010).

## 3. THE PIONEERS FROM HANS D. PFLUG TO RICHARD B. HOOVER

Nearly two decades later the problem of microbial fossils in carbonaceous meteorites was re-examined by Hans D. Pflug with special attention being paid to avoid the criticisms of earlier work (Pflug, 1984). Pflug used state-of-the-art equipment to prepare ultra-thin sections (< 1mm) of the Murchison meteorite in a contaminant free environment.

Thin slices of the Murchison meteorite were placed on membrane filters and exposed to hydrofluoric acid vapour. In this way *in situ* demineralisation was achieved, the mineral component being removed though the pores of the filter, leaving carbonaceous structures indigenous to the meteorite in tact. A wealth of morphologies with distinctive biological characteristics was thus revealed. Examples are shown in Figs 1, 2 and 3. Fig. 1 shows a rod-shaped bacterial shape, similar to structures found earlier in ocean sediments and in the atmosphere. Fig 2 shows structures uncannily similar to a well-known bacterium *pedomicrobium,* and Fig. 3 displays a clump of nanometric-sized particles with internal structure similar to a modern influenza virus. In view of the techniques used in the preparation of the slides, it could be asserted with confidence that all these structures are indigenous to the meteorite, not contaminants.

Microprobe analysis using laser mass spectroscopy, Raman spectroscopy, UV and IR spot spectroscopy were used to determine composition as well as to establish the indigenous nature of individual particles. Pflug's laser mass spectrum analysis on one of these particles is shown in Fig. 4. From Fig. 4, with many of the peaks yet to be unambiguously identified, we see that the particles with these biological-type morphologies also have chemical signatures fully consistent with degraded or fossilised microbial matter. Indeed comparisons of these structures with well-recognised microbial fossils in the Gunflint cherts showed nearly identical results in laser mass spectroscopy, demonstrating that the same organic functional groups were present in both situations. Further work by Pflug and Heinz (1997) confirmed these results and the criticism of contamination levelled against Claus and Nagy now became largely irrelevant. A more detailed re-appraisal of mass peaks in the data of Pflug and his collaborators is currently in progress (Wallis, Heinz and Wickramasinghe, 2010).

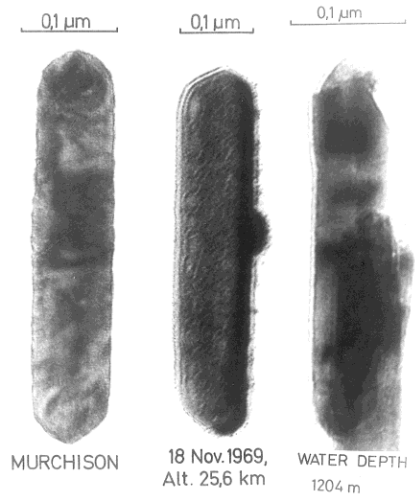

Figure 1. Figure captions are used to label the figure and help the reader understand the figure's significance. The caption should be centered underneath the figure and set in 9-point font. It is preferable for figures and tables to be placed at the top or bottom of the page.

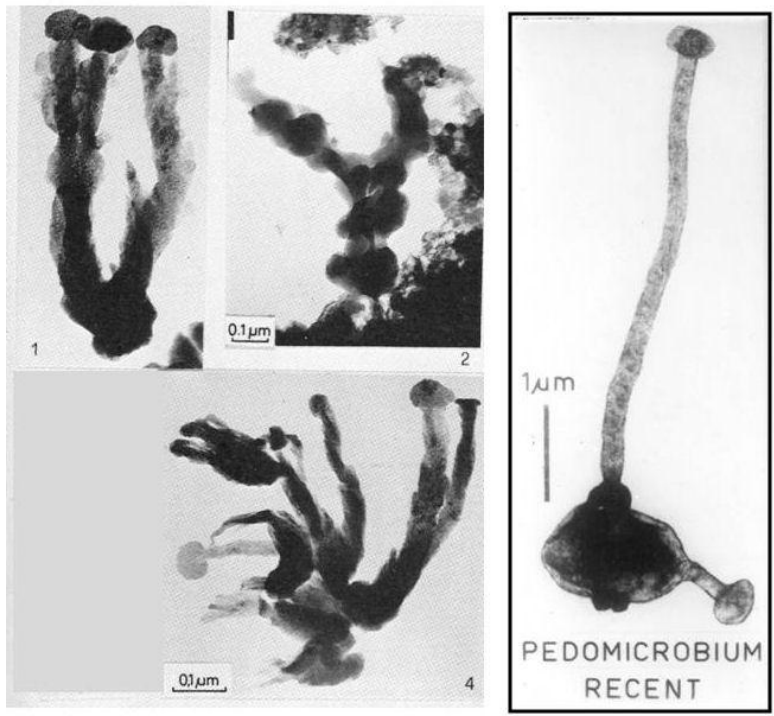

Figure.2 The comparison of a characteristically biological structure in the Murchison meteorite with a similar structure corresponding to a modern iron-oxidising microorganism – *pedomicrobium*.

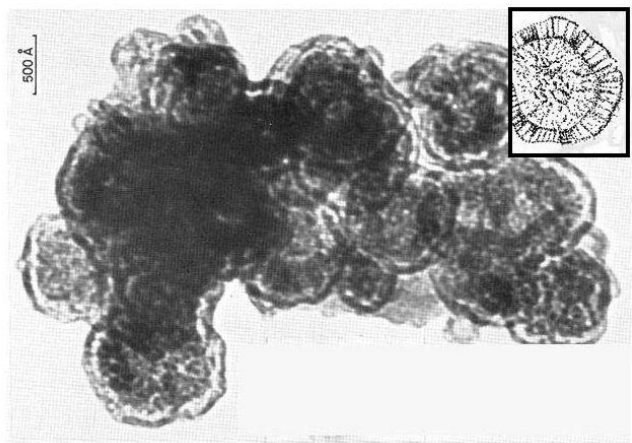

.

Figure 3  An electron micrograph of a structure resembling a clump of viruses – influenza virus – also found in the Murchison meteorite.  The drawing in the inset is a representation of a modern influenza virus displaying astounding similarities in structure to the putative clump of fossil viruses.

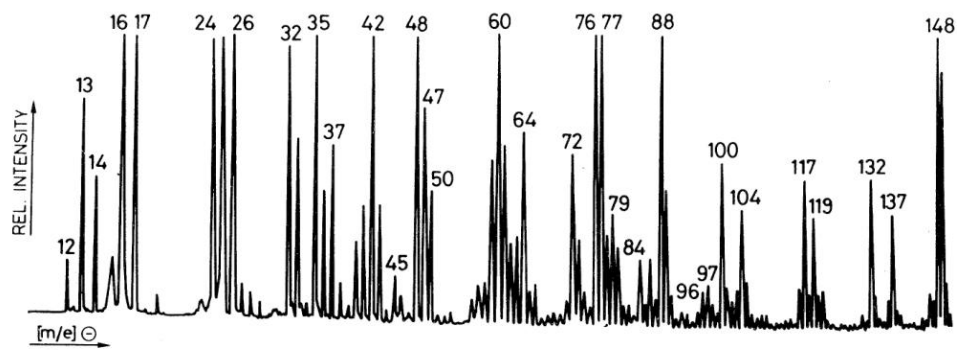

Laser mass spectrum (negative ions) from individual particle in thin section of the Murchison meteorite. Field of measurement ca. 1 μm. Attribution of signals: 12:$C^-$, 13:$CH^-$, 14:$CH_2^-$, 16:$O^-$, 17:$OH^-$, 19:$F^-$, 24:$C_2^-$, 25:$C_2H^-$, 26:$CN^-$, 28:$Si^-$, 32:$S^-$, 35:$Cl^-$, 36:$C_3^-$, 37:$C_3H^-$, 40–42, 45:fragmental carbonaceous groups, 48:$C_4^-$, 49:$C_4H^-$, 50:$C_4H_2^-$, 60:$SiO_2^-$ resp. $C_5^-$, 61:$C_5H^-$, 72:$C_6^-$, 73:$C_6H^-$, 76:$SiO_3^-$, 96:$C_8^-$, 97:$C_8H^-$, 108:$C_9^-$, 120:$C_{10}^-$, 121:$C_{10}H^-$. (Analysis and interpretations: H. J. Heinen, K.-D. Kupka).

Figure 4   Laser mass spectrum reproduced from Pflug (1984)

That debate over non-biotic artefacts resembling microbial fossils has persisted.  Recently a range of bio-indicators has been called up in evidence favouring biological origins of such structures as seen in Figs. 1-3 (Hoover 2006a). Likewise, claims for artefacts in the Martian meteorite ALH84001, have been dismissed as due to contamination or as non-biotic. However, McKay et al. (2009) recently reported studies of potential biofossils (carbon-associated "biomorphs„ ) in additional Martian meteorites Nakhla and Yamato-593.

Microfossils that confirm the pioneering work of Hans Pflug have in recent years been found in every carbonaceous chondrite studied by Hoover and co-workers in Russia (Moscow, Paleontological Inst) and USA (MSFC), but notably not in non-CC meteorites (Hoover 2006a,b).  They have used knowledge of microorganism morphology to tentatively identify some bio-fossils in CCs. Their ESEM and FESEM images of artifacts in the Murchison and Orgueil carbonaceous meteorites have uncovered filaments that typically exhibit dramatic chemical differentiation between the putative microfossil and the adjacent meteorite matrix.

Backscatter electron images at high resolution in Fig 5 shows a particularly impressive example comparing indigenous structures on a freshly cleaved surface of the Murchison meteorite with living cyanobacteria. Despite the quality of such data arguments still rage over the biogenecity of these features.

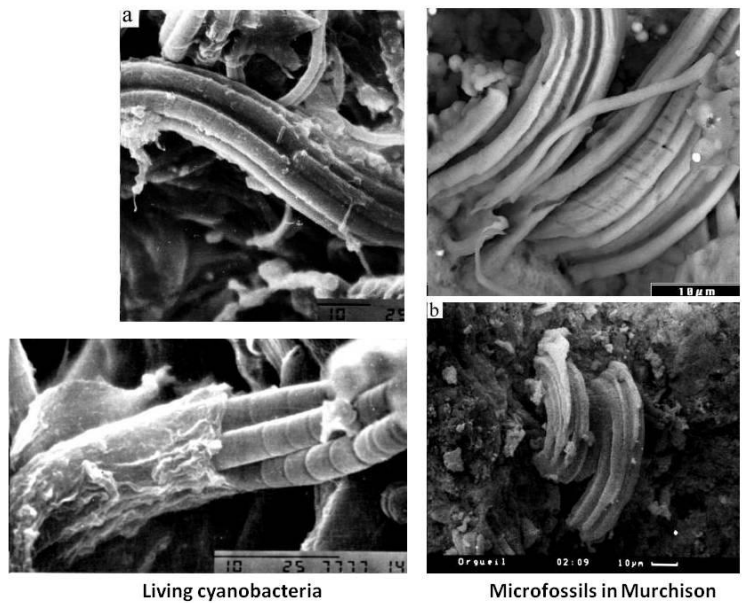

Figure 5  Structures in the Murchison meteorite (Hoover, 2005) compared with living cyanobacteria

The Cronin et al review (1988) asserted confidently that the 1960/70s argument over biogenic vs abiogenic origin had been resolved in favour of the latter, though they judged no particular mechanism was adequate to explain how such abiotic matter was created. However, as we have seen, recent studies indicate otherwise. Mukhopadhyay et al. (2009) believe the complex hydrocarbons may derive from bacteria and/or primitive algal remains (based on SEM-EDS, visual kerogen analysis using fluorescence, and white-light microscopy). Those authors found abundant alkanes (normal, cyclo-, and isoalkanes), alkyl aromatics, some polycyclic aromatic hydrocarbons, thiophenes, and nitrile compounds with biological signatures, especially within the Tagish Lake and Orgueil meteorites.

## 4. MICROFOSSILS IN COMETARY DUST

Carbonaceous chondrite (CC) meteorites such as the Murchison meteortite have 10% or more carbon, while cometary dust has a 30% CHON particle fraction (made of the light elements C,H,O,N) as well as mineral particles and particles of mixed composition, as discovered by the comet Halley space-probes.  IDPs recovered from the stratosphere are agglomerates that could be purely meteoritic or could have been processed in comets. Isotopically anomalous sub-micron components show the inclusion of pre-solar grains, while the silicate components indicate material condensed during an energetic phase of the early sun.

Traces of water can be detected in some samples and frequently the minerals in CC meteorites show evidence of aqueous alteration.  The identification of clay particles ejected from comet Tempel-1 by the Deep Impact probe in 2001 implied that aqueous alteration may indeed proceed in comets. The early IDP studies in the 1970s saw many of these particles as coming from comets; meanwhile, evidence for CC meteorites also originating from comets has grown. Enhanced IDP collection via balloon-borne cryosampling (Lal et al. 1996) and use of analytic techniques down to sub-micron scales have in recent years opened up further progress in studying fossils in both the mineral and the carbonaceous components of interplanetary dust particles.

Recovery of high velocity IDPs from the stratosphere  is effective because the Earth's tenuous air at ~ 100 km gradually slows down particles under 100μm while the smaller ones (<20μm) are only moderately heated (depending on density and zenith angle; Coulson and Wickramasinghe 2003), so lose their volatiles but retain more complex organics. They decelerated to low terminal velocities (some cm/s) and take weeks to months to descend below the stratosphere.

In the 1970s, high altitude flights (U2 aircraft) were used to collect them from the lower stratosphere, 18-20km altitude, on oiled sheets exposed outside aircraft flying at ~200m/s (Brownlee, 1978). This method suffered from the problem of contamination as well as breakage of the particles and a bias against small light ones (which tend to divert with the air stream). Moreover genuine interplanetary particles have to be diligently separated from terrestrial contaminants.

These so-called Brownlee particles, which were mostly in the form of fluffy aggregates of siliceous dust, have been found to contain extraterrestrial organic molecules, with a complexity and diversity approaching that recently reported for the Murchison meteorite (Clemett, et al, 1993). In a few instances microbial morphologies were discovered within individual Brownlee particles.

Figure 6 shows such a micron-sized carbonaceous structure in a Brownlee particle compared with a microbial fossil – an iron oxidising microorganism - found in the Gunflint cherts of N. Minnesota. The striking similarity seen here once again argues in favour of a biological rather than an abiotic explanation for the extraterrestrial particle (Hoyle, Wickramasinghe and Pflug, 1985; Bradley et al, 1984).

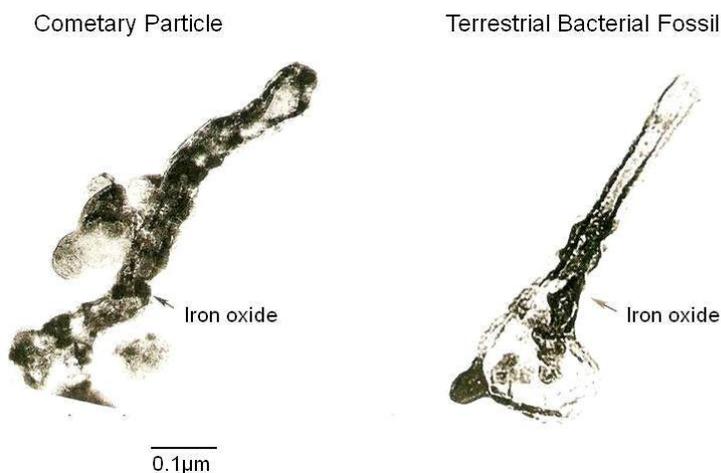

Figure 6: An organic particle in a Brownlee clump compared with a terrestrial bacterial fossil (Hoyle et al, 1985)

Cometary dust from comet Wild 2 collected in high velocity impacts with aerogel in the Stardust Mission would be expected to have a somewhat lower level of molecular diversity and complexity than the samples studied by Clemett et al (1993), and fragile biological structures would not have been recovered in tact. This is indeed borne out in examinations of Stardust material (Sandford et al, 2006).

## 5. CRYOPROBE SAMPLES OF UPPER STRATOSPHERE PARTICLES

Balloon flights launced by the Indian Space Research Organisation (ISRO) from the 1990s initially reached heights of ~30 km for smapling stratosphere CFCs, collecting frozen air in steel cylinders with all-metal valves (remotely controlled) immersed in liquid neon. The more recent flights reached heights of 40-45 km with all equipment ultra-clean and sterile to reduce contamination (Lal et al. 1996). In January 2001 this technique was used to collect pristine cometary dust aseptically using cryoprobes flown aboard balloons to heights of 41km (Harris et al, 2003; Narlikar et al, 2003; Wainwright et al, 2004).

Cometary dust particles were collected in the following manner: one set of samples is extracted from the cylinders by releasing the compressed air through micropore filters, another set is from filtering washings of particles adhering to the interior surface. At Cardiff we have used a few glass fibre filters, but mainly 0.45μm acetate filters. For better identifying the carbon fraction, some samples were transferred to silicon wafers (Miyake 2009).

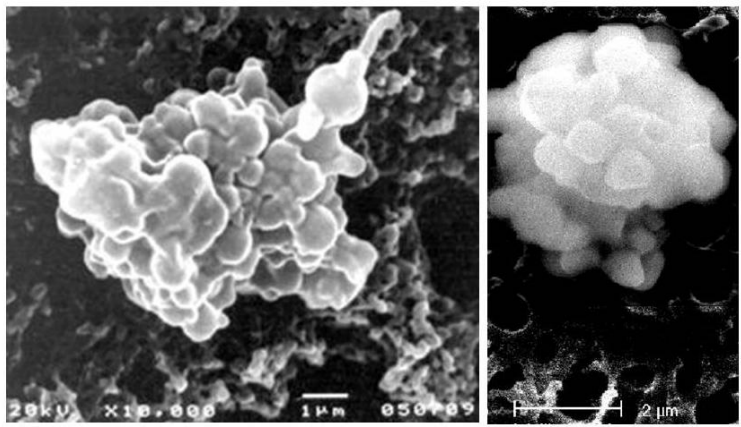

Figure 7  Putative microbial fossils in stratospheric aerosols (Harris, 2003; Wallis et al, 2006)

Amongst the aerosols collected were a rich harvest of pristine carbonaceous cometary dust particles bearing morphologies generally similar to bacterial fossils. Morphological similarities to cocoidal and rod-shaped bacteria have been noted by several investigators (Harris et al, 2003; Wainwright et al, 2008; Rauf et al, 2010). In a few instances evidence of fimbrae and biofilm appear to corroborate a biological interpretation, and in all cases shown here EDAX analyses have shown high C abundances (Wallis, et al, 2006; Wainwright et al, 2008). The height of 41 km from which the collections were made is arguably too high for lofting a 10μm sized clump of bacteria from the surface, so structures such as are seen in Fig.6 can be argued to represent infalling cometary dust.

Carbonaceous chondrite (CC) meteorites such as the Murchison meteortite have 10% or more carbon, while cometary dust has a 30% CHON particle fraction (made of the light elements C,H,O,N) as well as mineral particles and particles of mixed composition, as discovered by the comet Halley space-probes. IDPs recovered from the stratosphere are agglomerates that could be purely meteoritic or could have been processed in comets. Isotopically anomalous sub-micron components show the inclusion of pre-solar grains, while the silicate components indicate material condensed during an energetic phase of the early sun.

## 5.1  Identification of Acritarchs

Organic-walled microfossils found in terrestrial sedimentary rocks but of unidentified species are termed acritarchs. Acritarchs possess diverse shapes and forms and have been identified in pre-Cambrian sediments 3.2 By ago and are present in sediments of more recent times. Many of the specimens we have found in association with cometary dust collected in 2001, especially the ovoids, clearly resemble acritarchs.

Rossignol-Strick et al. (2005) reviewed the 1971 discovery of the acid-resistant, organic "hollow spheres" by Rossignol-Strick and Barghoorn (1971) and sought new examples in the Orgueil meteorite. Ovoid bodies in their new images were found to be composed of Fe-mineral within a thin carbonaceous sheath, like those tentatively identified as magnetite with ~0.2μm organic coatings (Alpern and Benkheiri 1973). By contrast, the organic globules found in the Tagish Lake meteorite (Nakamura et al. 2002) are mainly small structures - these μm-sized solid ovoids are quite distinct from the 1971 acid-resistant hollow spheres discovered by Rossignol-Strick and Barghoorn. The 5μm coccoid with 0.3μm carbonaceous envelope reported by Hoover (2006b) plus a similar one which Mukhopadhyay et al (2009) mapped in carbon and sulphur do, on the other hand, correspond to the 1971 discovery of acritarch-like structures in the Orgueil.

The Cardiff collection of IDPs contains many more acritarch examples. Those shown in Figs. 8-11 were found by the first SEM studies (Wallis et al. 2002; Miyake, 2009). The single spheres are spore-like, sometimes damaged (Fig. 8E) and often showing cracks (Figs.8 B, C, D, F). Cracks in the 'spores' are sometimes seen to widen under the microscope, but breaks in the surface appear to have existed pre-preparation. The surfaces show diverse structure and coatings – Fig. 8C shows partial coating, while the examples of Fig. 8 show thicker mineral deposits. Fibres (straight rods) about 0.5μm diameter are commonly attached (eg. Figs.8 A, C, E, F), while D shows finer whiskers embedded in the coating (Wallis et al. 2006).

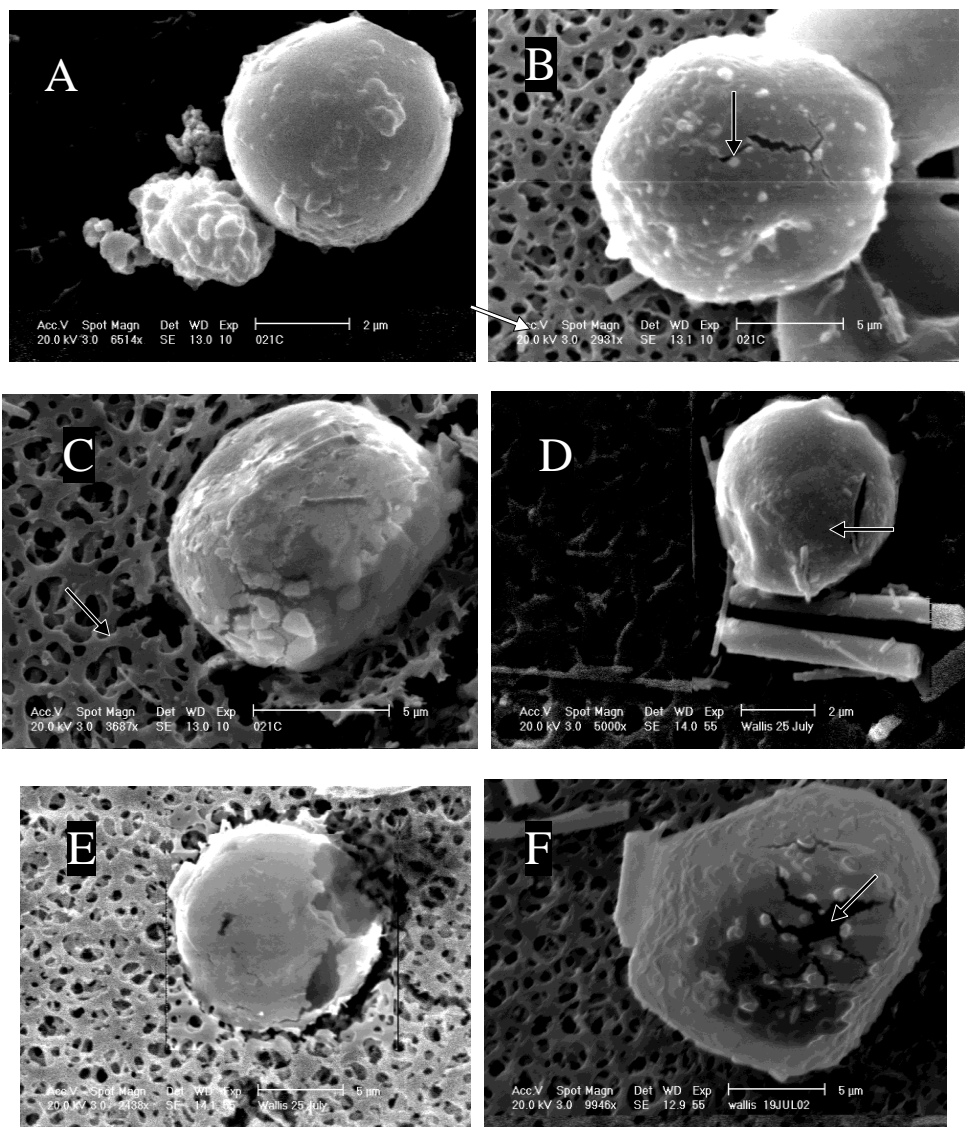

Figure 8 : Spherical IDPs resembling the Orgueil acritarchs from the Cardiff collection (Wallis et al. 2002). The samples are on a 0.45μm micropore filter of cellulose acetate stabilised by sputtered gold coatings, images by a Philips XL-20 scanning electron microscope (at 7 nanobar vacuum). Four specimens have cracks/slit (visibly widened or even caused by SEM exposure) while E has pieces missing, which show they are hollow shells. A: this 4μm-sized spherical particle is loosely attached to mineral IDPs; B, C and F are 10 μm spheres with cracks (black-head arrows) and disparate encrusted minerals. Particle B has a whisker attached to its underside (white-head arrow) while D and F's adjacent fibres would have separated on impact with the filter. D is a smaller acritarch with a slit rather than crack (arrowed).

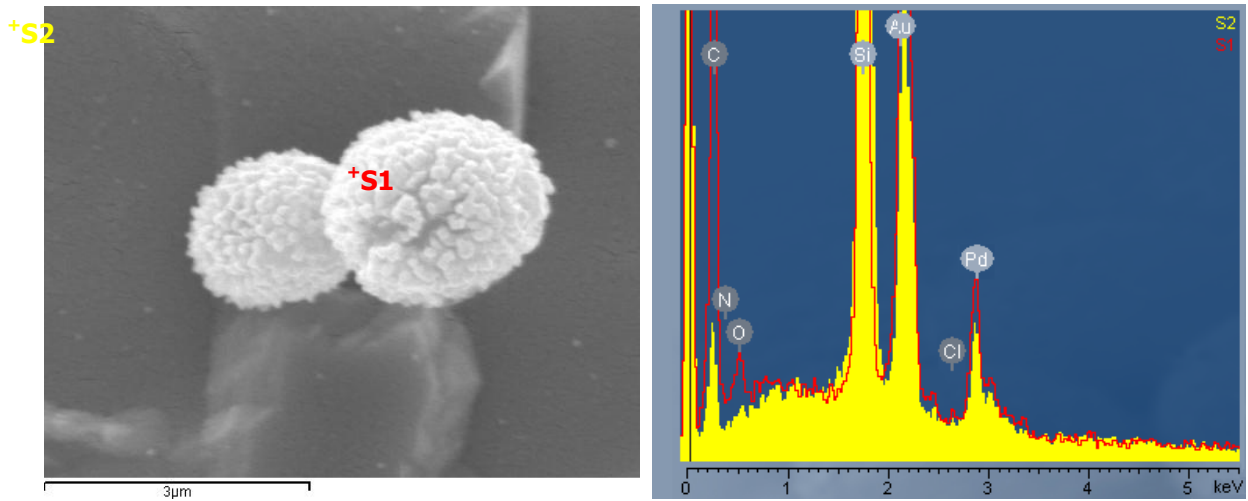

Figure 9: Further type of acritarch in the SEM studies (Miyake 2009) again after transfer to a silicon wafer. Specimen A shows cracking, indicative of a shell. The 2.5-4μm spheres are smaller than the acritarchs of Figs. 1-6 and have a distinctive surface structure. The EDX spectra refer to locations S1 and S2, colour-coded. S1 shows the particle is high in C (58%) and N (12%) but low in O (4%) and mineral elements apart from Si (uncertain due to the wafer; note the Pd is part of the gold coating, acting as a marker).

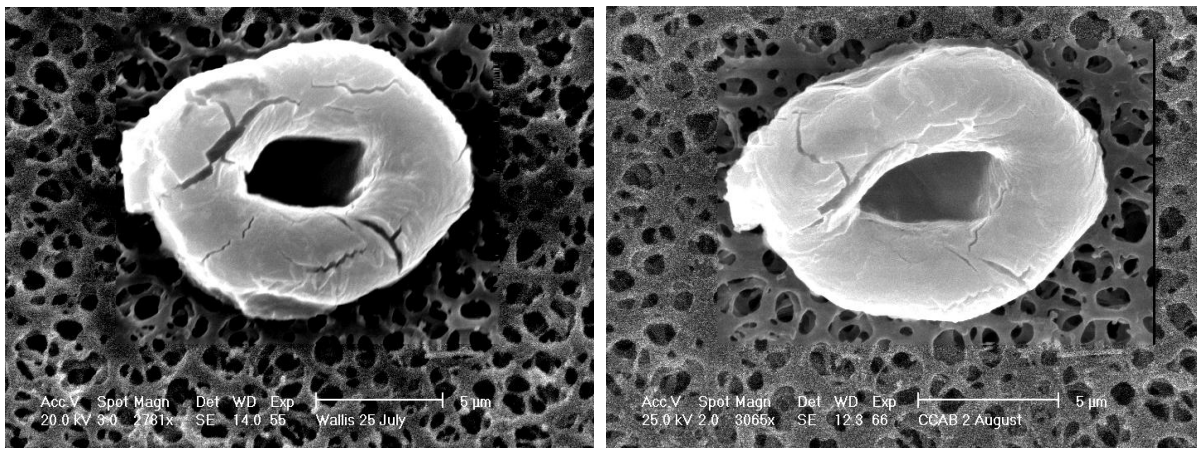

Figure 10: Further possible acritarchs showing a toroidal shell (the same particle under two angles of the electron beam) which would be a novel type of acritarch (also found in the Tagish Lake CC. Rauf et al., 2010a).

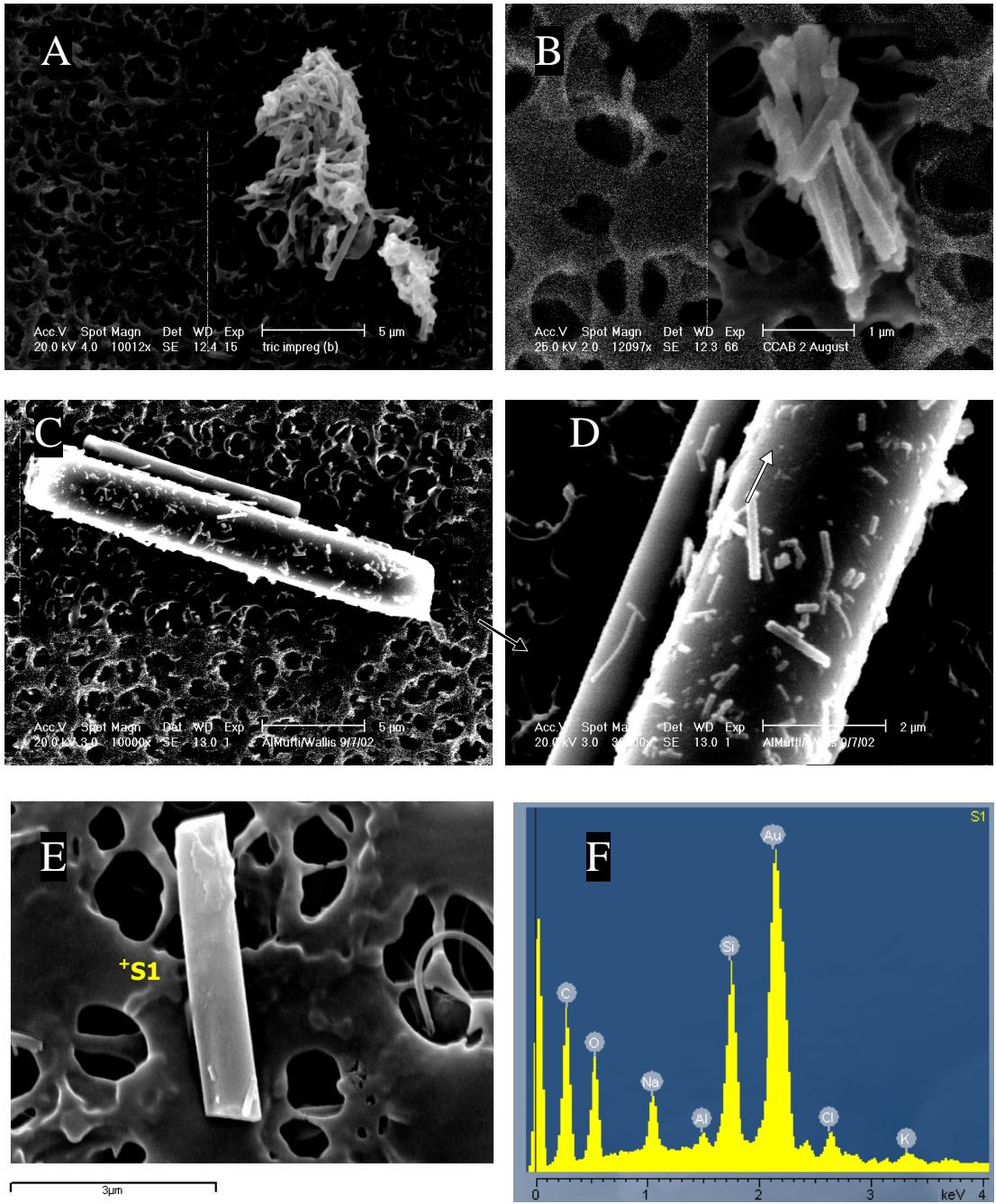

Figure 11 Examples of individual siliceous fibres. A shows three fibres originally stuck together, whereas E and B are separate. C is a large (3µm diameter x 20µm long) fibre with 'baby'. D gives the magnified centre of C showing sub-micron whiskers in the surface: eg. a short 200-300nm one (white-head arrow) and a long 1.5 µm one (black-head arrow). Similar whiskers are also evident on the surface of B. The spectrum F of fibre E shows the main Si peak with some Na, K and Cl (C appears strong but has an uncertain contribution from the acetate background).

Silceous fibres were also found to be common in our recovered acritarchs as isolated rods or attached to other grains. Initially we assumed these to be terrestrial contaminants (glass fibres). However, fibres are both attached or associated with acritarchs, as seen above, and embedded in loose aggregates, as shown in Fig.11.

We have discounted a possible astrophysical origin of the siliceous fibres in outflows from the sun and stars (Miyake et al. 2009). Asteroidal and cometary silicates are understood as crystalline or fine phyllosilicate clays identified by infra-red emissions in comet Tempel-1's dust ejected by the Deep Impact collision (Lisse *et al.*, 2006). The siliceous rods and fibres fit with neither origin. The proposal of comets as a potential habitat for siliceous diatoms dates to 1985 (Hoover et al. 1986) because of IR spectral similarities and because polar diatoms survive in polar ice at low light levels, hibernating at low temperatures.

Some marine diatoms possess whiskers (pili = hair-like extensions), others have intricate siliceous exoskeletons. Figure 13 shows several examples of living diatom exhibiting siliceous whiskers.

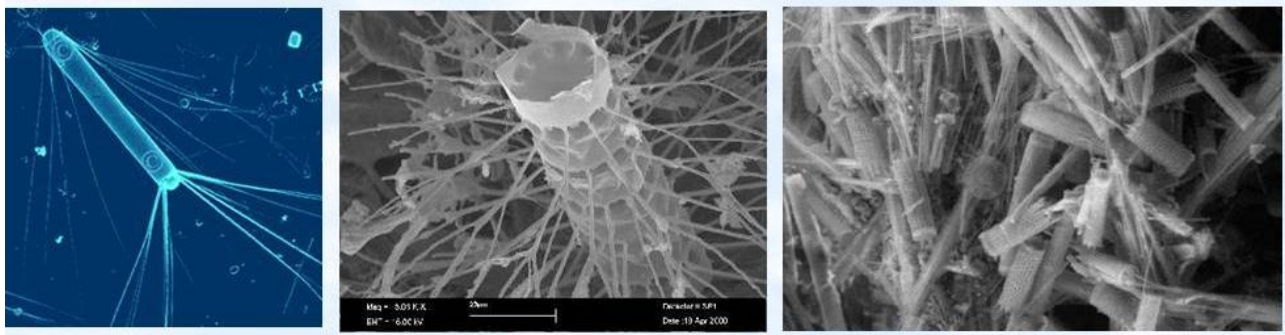

Figure 13 *Left* – Image of living antarctic diatom forming several whiskers at the both ends of body, x200 (© Breger of Lamont-Doherty Earth Observatory of Columbia University). *Centre* – Image of branched marine diatom collected at Rockey point, Mexico, scale bar - 25µm (© Boyer of Northen Arizona University). *Right* – deposit at a bottom of lake showing a range of sizes of siliceous diatom fragments (© Wetmore of University of California Berkeley Museum of Paleontology).

From comparisons of Figs 11, 12 with Fig. 13 we hypothesise that our siliceous fibres and whiskers are fragments of fossil diatoms that lived on comets and/or icy satellites. If embedded in ice ejected via collisions, they would be freed as the ice sublimates away. If on the surface of a comet (or in Martian seasonal ice deposits) the fibres would attach to other minerals as the ice sublimates away, becoming embedded in the loose aggregates.

Acritarchs when found in terrestrial sedimentary rocks are presumed to be biological, eukaryotes, but of undetermined species. The discovery in the Orgueil meteorite were explained as artifacts. Our IDP acritarchs would challenge such abiotic explanations. Though both may derive from comets, those aggregated in carbonaceous chondrites have undergone some compaction and aqueous alteration, while the IDP examples may have been released within icy matrices. The IDP samples are less altered geochemically than Orgueil's; they have mineral coatings and are just loosely associated with other materials considered to be of cometary origin. Thus, if C-isotope studies (via eg. an ion microprobe) confirm their space origin, the specimens of Fig. 10 provide a strong indicator of cometary life.

## 6. VIABLE BIOLOGICAL STRUCTURES IN COMET DUST

In addition to the profusion of acritarchs discussed in section 5, interesting, although disputed, evidence of a collection of viable microbes were also discovered. Fig 14 shows a clump of cocci from the 2001 collection flourescing under the application of a carbocyanine dye. This confocal microscope image shows evidence of a viable but not culturable microorganism. Fig. 15 shows a collection of other unpublished confocal images fluorescing in a dye that detects DNA. This work was carried out in Cardiff by Melanie Harris (Harris et al, 2003).

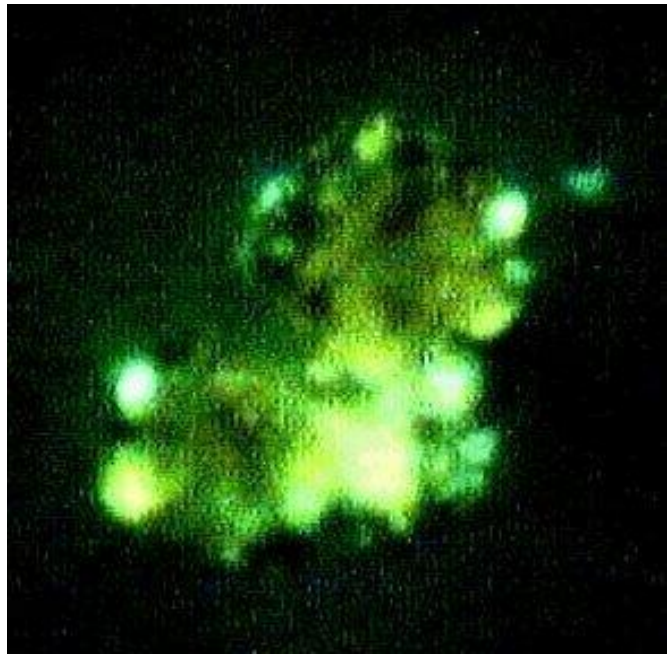

Figure 14 Clump of putative cells fluorescing under confocal microscopy under application of carbocyanine dye

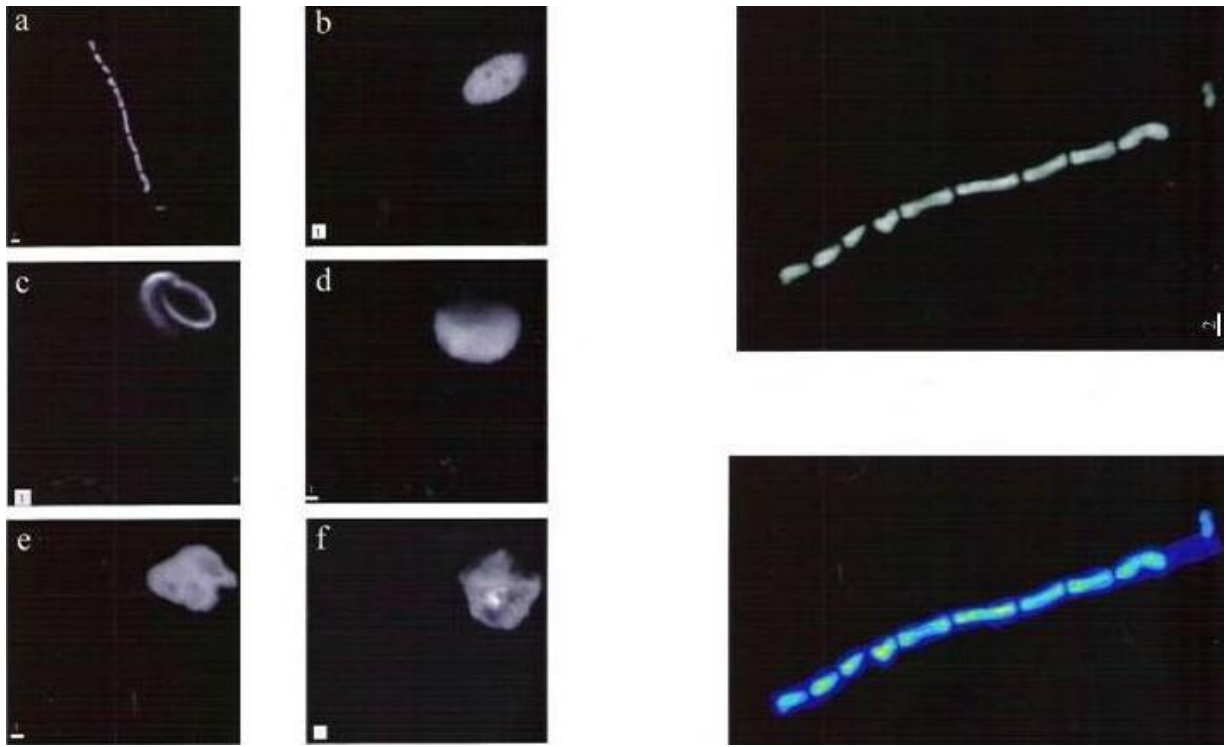

Figure 15: Confocal images of 0.7µm glass microfibre membranes stained with the nucleic acid dye acridine orange. Portions of the filters were isolated from samples collected at 41km. (Courtesy Melanie Harris)

# 5. CONCLUDING REMARKS

In view of the recent revival of interest I the theory of cometary panspermia (Wickramasinghe, 2010) a careful re-assessment of all its predictions now appears timely. Cometary panspermia argues that comets are the carriers and distributors of cosmic life as well as the sites of replication of cosmic bacteria. Since carbonaceous chondrites may be regarded as being the end product of the evolution of comets, progressively denuded of volatiles after very many perihelion passages, such objects would naturally be expected to carry chemical and morphological signatures of life.

Early reports in the 1960's purporting to find microfossils (organised elements) in meteorites came to be quickly discounted after a few instances of contamination by ragweed pollen contamination were discovered. This was an unfortunate development because later attempts to re-open the subject of microfossils in meteorites became inevitably tainted with prejudice. In this article a sample of the most convincing results obtained by Hans Pflug, Richard Hoover and ourselves are assembled. From the best evidence available it could be seen that the presence of microfossils in carbonaceous meteorites is strong – even compelling.

Since comets in our view are the repositories, incubators and transporters of cosmic life, this work provides strong support for the theory of cometary panspermia. In our view life most likely originated on a scale that far transcends the size of the galaxy or even the local cluster of galaxies. We have argued elsewhere that life is indeed a cosmological phenomenon. In the HGD cosmology discussed by one of us (CH) the best sites for the origin of life is within primordial planets that began their condensation process after the plasma to neutral transition some 300.000yr after the Big Bang. In another paper in this symposium we discuss how these sites began to evolve in radioactively-melted cores of icy planets 30My after the Big Bang. Living forms of the shapes shown in the present paper, together with a profusion of organic of the kind discovered by Schmitt-Koplin et al (2010) were locked in planets whose CNO content was estimated to be $\sim 10^{27}$ g. The ingress of a single such planet into the protosolar nebula would provide the material for $10^{11}$ comets known to be present in our solar system. The evidence of a disintegrating planet in the Helix Nebula shown in Fig 16 provides striking evidence of such a process in action.

## Spitzer infrared view of nearby Helix planetary nebula

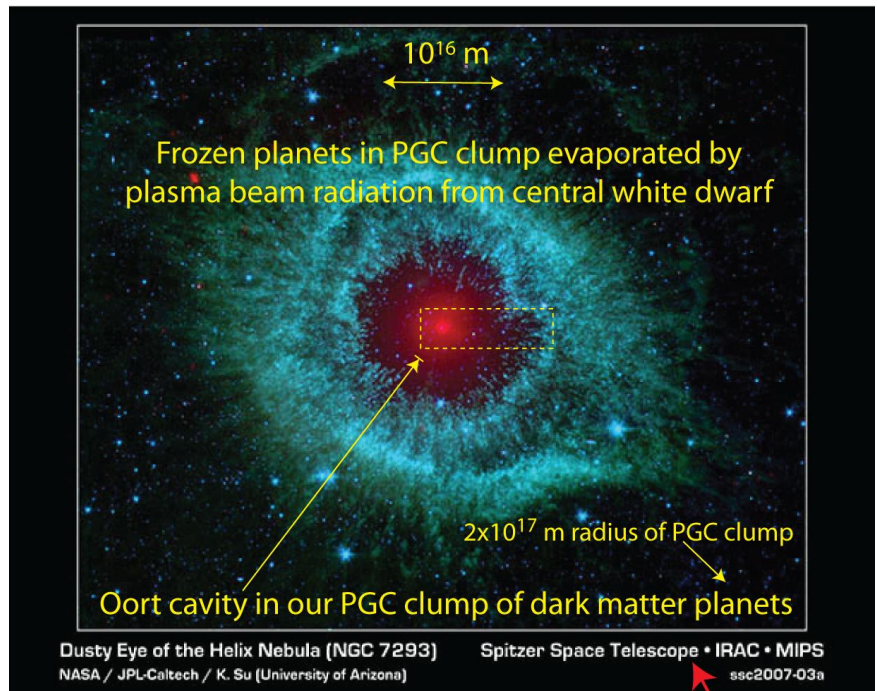

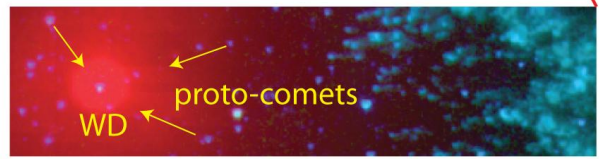

Figure 16. Spitzer space satellite infrared view of Helix Planetary Nebula. Approximately 40,000 primordial planets are detected, partially evaporated in the PGC clump of a trillion by plasma beam radiation from the central white dwarf fed by proto-comets drawn by gravity toward the center from the boundary of the Oort cavity. Dust from the evaporated comets is shown in red. H and He gas from the planets is converted to carbon by the white dwarf until it exceeds the Chandrasekhar limit of 1.44 solar masses and explodes as a Supernova Ia event.